\documentclass[10pt]{article}
\usepackage[a4paper,margin=3.5cm]{geometry}
\usepackage{graphicx}
\usepackage{authblk}
\usepackage{amsmath, amssymb}
\usepackage{hyperref}
\usepackage{float}
\usepackage{caption}
\usepackage{subcaption}
\usepackage{cite}
\usepackage{booktabs} 
\usepackage{multirow}
\usepackage{comment}

\title{\bf A depth-dependent, transverse shift-invariant operator for fast iterative 3D photoacoustic tomography in planar geometry}

\author[1]{Ege Küçükkömürcü}
\author[2]{Simon Labouesse}
\author[1]{Marc Allain}
\author[1*]{Thomas Chaigne}

\affil[1]{\small Aix Marseille Univ, CNRS, Centrale Marseille, Institut Fresnel, Marseille, France}
\affil[2]{\small Rimeo, Montpellier, France}
\affil[*]{\small Corresponding author: thomas.chaigne@fresnel.fr}

\date{}

\begin{document}
\maketitle

\begin{abstract}
Iterative model-based image reconstruction in photoacoustic tomography (PAT) enables principled incorporation of detector physics, object-related priors, and complex acquisition strategies. However, for three-dimensional (3D) imaging scenario, the computational cost is often dominated by repeatedly solving wave equations. We propose a fast forward model for planar detection geometries that exploits transverse shift invariance. This symmetry enables to compute the full acoustic field from a 3D object, as a result of a set of 2D convolutions with depth-dependent impulse responses. This formulation yields a FFT-based forward operator and its corresponding discrete adjoint operator, making iterative reconstruction faster without calling partial differential equation (PDE) solvers at each iteration. We validate the model against commonly used PDE solver under matched discretization and boundary settings, and demonstrate accelerations of up to 2 orders of magnitude for iterative reconstructions from experimental all-optical photoacoustic datasets.
\end{abstract}

\section{Introduction}
Photoacoustic tomography (PAT) is a hybrid imaging modality that combines the high optical contrast of light absorption with the high spatial resolution deep inside soft tissue of ultrasound detection \cite{beard_biomedical_2011,wang_photoacoustic_2009}. A nanosecond-pulsed laser excitation induces localized thermoelastic expansion, generating broadband ultrasound waves that propagate through soft biological tissue and can be recorded at the tissue surface. Acoustic inversion consists of recovering the initial pressure distribution from these measured time series. This image reconstruction is an inverse problem governed by the acoustic wave equation and, in practical settings, is ill-posed due to limited aperture, specific detector response, and measurement noise. As a result, accurate and computationally efficient reconstruction remains a central challenge, particularly in three-dimensional (3D) imaging.

Classical reconstruction methods in PAT include analytic and direct inversion approaches derived from explicit inversion formulas of the spherical-mean operator. These include filtered backprojection variants \cite{finch_inversion_2007,haltmeier_filtered_2005}, the universal back-projection formula \cite{xu_universal_2005}, and extensions tailored to specific acquisition geometries \cite{burgholzer_exact_2007,hristova_reconstruction_2008}. When their underlying assumptions are closely met (e.g.\ full measurement aperture, homogeneous acoustic properties, and ideal detector responses), these methods can provide qualitative reconstructions with low computational cost. In experimental systems, however, these assumptions are often violated (e.g.\ limited-view, finite bandwidth, and non-ideal sensor responses), which can lead to pronounced artefacts and reduced robustness to noise \cite{frikel_artifacts_2015,zhu_mitigating_2023,rietberg_artifacts_2025}. Beyond analytic backprojection-type formulas, several direct inversion methods have also been derived in the frequency domain, where reconstruction can be implemented efficiently using FFTs \cite{kostli_two-dimensional_2003,kostli_temporal_2001}. In planar detection geometry in particular, the measured data can be transformed over the sensor coordinates and time, related to the spatial spectrum of the initial pressure through the acoustic dispersion relation, and inverted by an inverse Fourier transform, yielding fast 3D reconstructions. Such FFT-based approaches are attractive because of their computational efficiency and have become a standard direct baseline in planar configurations. Nevertheless, in practical systems it remains sensitive to limited-view geometry, finite detector bandwidth, non-ideal sensor responses, and measurement noise, which can lead to reconstruction artefacts.

A more general physics-based alternative is time-reversal (TR) reconstruction, in which the measured wave field is numerically propagated backward in time \cite{fink_time_1992,treeby_photoacoustic_2010}. TR naturally accommodates more complex sensor geometries \cite{treeby_k-wave_2010} and, when available, heterogeneous acoustic models \cite{cox_artifact_2010}. Widely used k-space solvers such as \texttt{k-Wave} \cite{treeby_k-wave_2010} provide practical implementations that can be adapted to a variety of experimental imaging systems. Because TR reconstructs an image through a single numerical wave-equation solve, it is often used as a direct physics-based baseline. Yet it suffers from other limitations previously mentioned.

Model-based (MB) reconstruction has been introduced to mitigate these effects by formulating photoacoustic image reconstruction as a regularized iterative optimization problem. A forward model first predicts the acoustic signals generated by a current estimate of the initial pressure distribution, and these predicted data are compared with the experimentally measured time series through a data-fidelity term, often combined with regularization and constraints. The image estimate is then updated iteratively to reduce this discrepancy, using the adjoint operator to map the data residual back into the image domain. This general framework can be solved using a variety of iterative optimization schemes, including first-order methods such as gradient descent \cite{nocedal_numerical_1999}, proximal algorithms \cite{boyd_proximal_2014}, accelerated methods such as FISTA \cite{beck_fast_2009}, and primal--dual algorithms \cite{chambolle_first-order_2011,condat_primaldual_2013}. In all cases, these methods require repeated applications of the forward operator and its adjoint. In 3D PAT, these operators are commonly implemented through full wave-equation propagation using finite-difference time-domain or k-space solvers \cite{arridge_adjoint_2016}. Although many minimization schemes can converge in relatively few iterations \cite{arridge_accelerated_2016}, the overall runtime is often dominated by the cost of these repeated wave-equation solves \cite{lucka_enhancing_2018}. For example, reconstructing a $1\,\mathrm{cm}^3$ volume with $100\,\mu\mathrm{m}$ isotropic voxel size corresponds to a total of $N_{xyz}=10^6$ voxels. For a typical acquisition duration of a few microseconds, corresponding to $N_t\sim 1\times 10^2$--$10^3$ time steps, an MB reconstruction with typically $10$ to $50$ iterations and one forward and one adjoint evaluation per iteration entails approximately $20$ to $100$ full 3D wave-equation solves, which dominate the computational cost. The high computational burden associated with repeatedly solving the acoustic wave equation in iterative reconstruction has motivated significant effort toward alternative forward-model formulations that would reduce runtime while preserving physical fidelity. Common strategies include analytic or semi-analytic representations, precomputed impulse responses, and exploiting geometric structure to avoid full 3D wave propagation during each iteration.
Several works have explored alternatives to full time-domain wave-equation solvers, including analytic or semi-analytic reformulations of the photoacoustic forward model based on the spherical-mean operator \cite{boink_framework_2018}, k-space-derived models \cite{buehler_modelbased_2011,rosenthal_modelbased_2011}, and explicit system-matrix approaches built from synthetic or experimentally measured spatio-temporal impulse responses \cite{araque_caballero_model-based_2014,queiros_modeling_2013,chowdhury_synthetic_2020,dean-ben_practical_2022}. Once constructed, such models provide accurate matched forward and adjoint operators and avoid re-solving the wave equation at each iteration. However, unless additional structure or compression is exploited, they generally require storing voxel-dependent spatio-temporal responses, so both memory usage and operator application scale poorly in 3D. For a volume of $N_{xyz}$ voxels, storing voxel-wise responses over $N_t$ time samples scales as $\mathcal{O}(N_{xyz}N_t)$. Even considering reasonable grid size (e.g.\ $N_x,N_y,N_z\gtrsim 256$ and $N_t\sim 10^3$), this quickly reaches tens to hundreds of gigabytes, limiting scalability.
To address this, several authors have exploited symmetry and invariance properties of the acquisition geometry to compress the operator representation. In spherical or hemispherical detection geometries, the impulse response depends primarily on the source--detector distance, enabling rotational or radial invariance to be leveraged for compact representations \cite{araque_caballero_model-based_2014}. Such symmetry-based compression has proven highly effective for curved-array tomography. However, these approaches are intrinsically tied to non-planar detection geometries and do not directly extend to planar sensor configurations, where transverse shift invariance can be exploited instead.

In this work, we propose a structured forward model for 3D photoacoustic imaging with planar detection. In a homogeneous medium, acoustic propagation from a source located at a given depth to the sensor plane is transversely shift invariant. The forward model can therefore be expressed as a superposition of depth-dependent 2D convolutions in the transverse plane. Using depth-indexed spatio-temporal impulse responses, we implement matched forward and adjoint operators as batched 2D FFT-based linear convolutions, eliminating PDE solvers during iterations and substantially reducing the per-iteration cost of MB solvers by more than two orders of magnitude.

\section{Theory}
We consider photoacoustic signal generation and propagation in a homogeneous,
isotropic, and lossless acoustic medium with uniform sound speed $c$.
Under conditions of thermal and stress confinement, the acoustic pressure field
$p(\mathbf r,t)$ satisfies the homogeneous wave equation~\cite{wang_photoacoustic_2009}
\begin{equation}
\left(
\nabla^2 - \frac{1}{c^2}\frac{\partial^2}{\partial t^2}
\right) p(\mathbf r,t) = 0,
\label{eq:wave_equation}
\end{equation}
with initial conditions
\begin{equation}
p(\mathbf r,0) = p_0(\mathbf r)=\Gamma\mu_a(\mathbf r)\Phi(\mathbf r),
\qquad
\left.\frac{\partial p(\mathbf r,t)}{\partial t}\right|_{t=0} = 0,
\label{eq:initial_conditions}
\end{equation}
where $p_0(\mathbf r)$ is the initial pressure distribution, $\Gamma$ is the Gr\"uneisen parameter,
$\mu_a$ is the optical absorption coefficient, and $\Phi$ is the optical fluence.
In this work we address the \emph{acoustic} inverse problem and treat $p_0$ as the unknown
to be reconstructed. We start of by assuming a planar acquisition geometry with a sensor plane located at $z_s=0$ .
We further assume identical, spatially invariant impulse response (e.g., bandwidth and directivity)
for all sensor elements. Our framework allows for this response to be incorporated into the spatio-temporal kernels
defined below (for instance using temporal convolution).

\subsection{Depth-dependent transverse shift invariance}
We start of by writing the spatial coordinate as $\mathbf r=(\mathbf r_\perp,z)$ with $\mathbf r_\perp=(x,y)$ and
a sensor location on the plane as $\mathbf r_s=(\mathbf r_{s,\perp},0)$. For a lossless homogeneous medium,
the Green's function representation implies that the measured pressure at $\mathbf r_s$ can be written
as~\cite{wang_photoacoustic_2009}
\begin{equation}
p(\mathbf r_s,t)
=
\frac{\partial}{\partial t}
\int_{\mathbb R^3}
\frac{
\delta\!\left(t-\|\mathbf r-\mathbf r_s\|/c\right)
}{
4\pi\|\mathbf r-\mathbf r_s\|
}
\,p_0(\mathbf r)\,
\mathrm d\mathbf r,
\label{eq:greens_solution}
\end{equation}
i.e., by superposition of time-shifted spherical waves whose spatial dependence is only through the
relative displacement $\mathbf r-\mathbf r_s$.

Using $\mathrm d\mathbf r = \mathrm d\mathbf r_\perp\,\mathrm dz$ and restricting the object support to $z>0$,
\eqref{eq:greens_solution} becomes
\begin{equation}
p(\mathbf r_{s,\perp},t)
=
\frac{\partial}{\partial t}
\int_{0}^{\infty}\!
\int_{\mathbb R^2}
\frac{
\delta\!\left(t-\sqrt{\|\mathbf r_{\perp}-\mathbf r_{s,\perp}\|^2+z^2}/c\right)
}{
4\pi\sqrt{\|\mathbf r_{\perp}-\mathbf r_{s,\perp}\|^2+z^2}
}
\,p_0(\mathbf r_\perp,z)\,
\mathrm d\mathbf r_\perp\,\mathrm dz.
\label{eq:forward_plane_explicit}
\end{equation}

Crucially, the transverse coordinates appear only through the difference
$\mathbf r_\perp-\mathbf r_{s,\perp}$.
For a fixed depth $z$, we therefore define the depth-indexed spatio-temporal impulse response on the
sensor plane
\begin{equation}
h_z(\mathbf r_\perp,t)
\triangleq
\frac{\partial}{\partial t}
\left[
\frac{1}{4\pi \sqrt{\|\mathbf r_{\perp}\|^2+z^2}}
\,
\delta\!\left(
t-\frac{\sqrt{\|\mathbf r_{\perp}\|^2+z^2}}{c}
\right)
\right].
\label{eq:hz_def}
\end{equation}

\begin{figure}[t]
  \centering
  \includegraphics[width=0.75\linewidth]{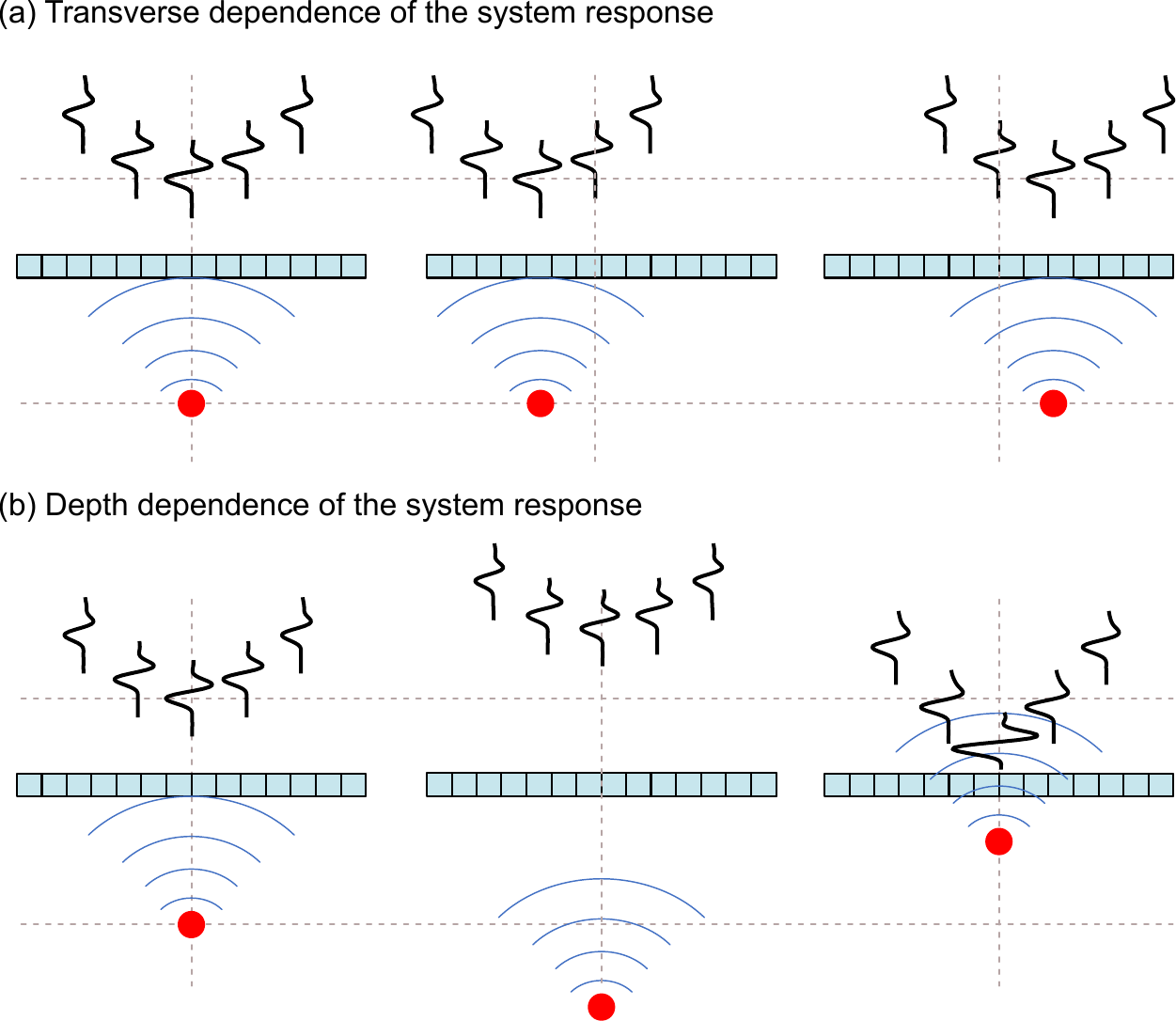}
    \caption{\textbf{Depth-dependent transverse shift invariance in planar photoacoustic acquisition.}
    (a) Transverse shift invariance: at fixed depth $z$, laterally translating a point absorber
    translates the measured wave-field by the same amount, implying convolution in $(x,y)$.
    (b) Depth dependence: increasing $z$ increases time-of-flight and changes the
    wavefront curvature on the sensor plane. These effects motivate the depth-indexed convolutional model
    \eqref{eq:conv_model_continuous}--\eqref{eq:discrete_forward}.}
  \label{fig:shift_invariance_depth}
\end{figure}

Then \eqref{eq:forward_plane_explicit} can be written as an integral over depths of \emph{2D transverse
linear convolutions}:
\begin{equation}
p(\mathbf r_{s,\perp},t)
=
\int_{0}^{\infty}
\Bigl[
(h_z(\cdot,t) *_{\perp} p_0(\cdot,z))(\mathbf r_{s,\perp})
\Bigr]
\,\mathrm dz,
\label{eq:conv_model_continuous}
\end{equation}
where $*_{\perp}$ denotes 2D convolution over $(x,y)$. As illustrated in Figure~\ref{fig:shift_invariance_depth}, Equation~\eqref{eq:forward_plane_explicit} expresses \emph{depth-dependent transverse shift invariance}.
For each fixed depth $z$, translating $p_0(\mathbf r_\perp,z)$ by $\Delta\mathbf r_\perp$ translates by the same amount the measured
field on the sensor plane. Depth influences the measurements through time-of-flight and geometric spreading, which are encoded in the depth-dependent impulse responses $h_z$.
This leads to the 2D convolution structure in the $(x,y)$ transverse plane (Equation~\eqref{eq:conv_model_continuous}).

In practice, the depth-dependent impulses responses $\{h_z\}_{z>0}$ can be obtained either from the analytic expression
\eqref{eq:hz_def} or, as in this work, by numerically simulating the response to unit-amplitude point
absorbers placed at successive depths using k-Wave~\cite{treeby_k-wave_2010}. This underlying transverse shift invariance does \emph{not} contradict the spatially varying reconstruction quality that is commonly observed in the field \cite{hauptmann_model-based_2018-1,choi_practical_2020,chowdhury_synthetic_2020}. This effect solely originates from limited-view detection geometries, rather than via a transverse dependence of the propagation model itself. It only truncates the translated wave-fields, hence yielding different reconstructed images of point sources at different locations in the volumes.

\subsection{Discrete convolutional forward and adjoint operators}
Let us discretize the reconstruction volume over a regular Cartesian
grid $\mathcal{G}_\rho = \mathcal{G}_{xy} \times \mathcal{G}_z$ with
$$
\mathcal{G}_{xy} = \{ (x_n,y_m)\, : \, n = 1\cdots N_x,\, m=1\cdots N_y\}
\quad \text{and} \quad \mathcal{G}_z = \{ z_n \}_{n=1}^{N_z}.
$$
The measurement space is also discretized and the dataset is defined over another   
Cartesian (spatio-temporal)  mesh $\mathcal{G}_d =\mathcal{G}_{xy} \times \mathcal{T}$ where $\mathcal{T} = \{ t_n \}_{n=1}^{N_t}$ is the
discretized time dimension.

Introducing the following notations for the quantity to
reconstruct and the sensor data, respectively
$$
\begin{array}{rclr}
  \rho (\mathbf{r}) &\equiv& p_0(\mathbf{r}), &\quad  \mathbf{r}\in \mathcal{G}_\rho\\
  d (\mathbf{r}_\perp, t) &\equiv& p(\mathbf{r}_\perp, t) , &\quad  (\mathbf{r}_\perp,t) \in \mathcal{G}_d,  
\end{array}
$$
the depth integral in~\eqref{eq:conv_model_continuous} can be
approximated by the Riemann sum
\begin{equation}
d (\mathbf{r}_\perp, t) =
\sum_{z\in \mathcal{G}_z} h_z(\cdot, t)
\otimes_{\perp} \rho(\cdot,z),
\qquad t\in\mathcal{T},
\label{eq:discrete_forward}
\end{equation}
where '$\otimes_{\perp}$' denotes the convolution over the 2D mesh $\mathcal{G}_{xy}$.
For notational simplicity, we stress that constant discretization
factors (e.g., sampling step-size along any direction) in
\eqref{eq:discrete_forward} are  absorbed in the depth-dependent
convolution kernel $h_z$. The relation above defines a linear
mapping 
\begin{equation}
d = H\rho,
\label{eq:H_operator}
\end{equation}
with $H:\, \mathbb R^{N_x\times N_y\times N_z}\, \longrightarrow \, \mathbb
R^{N_x\times N_y\times N_t}$ a matrix  whose action performs a sum of transverse
convolutions with depth-dependent kernels.
The corresponding \textit{adjoint} operator, denoted  $H^*$, can then
be defined as the unique linear mapping satisfying
\begin{equation}
\langle H\rho,\,d\rangle_{D} = \langle \rho,\,H^* d\rangle_{O}
\label{eq:adjoint_def_clean}
\end{equation}
with $\langle \cdot, \cdot \rangle_D$ and $\langle \cdot, \cdot
\rangle_O$ a \textit{scalar-product} for the data
(i.e., measurement) space and the object (i.e., initial pressure)
space, defined as
\begin{equation}
  \langle d_1,d_2\rangle_{D} \triangleq \sum_{\mathbf{r}_\perp \in \mathcal{G}_{xy}} \sum_{t\in \mathcal{T}}
  d_1(\mathbf{r}_\perp , t) \, d_2(\mathbf{r}_\perp , t) 
\quad \text{and} \quad
  \langle \rho_1,\rho_2\rangle_{O} \triangleq \sum_{\mathbf{r} \in \mathcal{G}_{\rho}} 
  \rho_1(\mathbf{r}) \, \rho_2(\mathbf{r}) 
\label{eq:euclidean_inner_products_clean}
\end{equation}
We can then establish from \eqref{eq:adjoint_def_clean} that
the adjoint $H^*$ is performing a sum of 2D correlation with the
convolution kernels
\begin{equation}
(H^* d)(\mathbf{r}_\perp, z) = \sum_{t\in \mathcal{T}} h_z(\cdot,t)\star_\perp d(\cdot,t),
\label{eq:adjoint_final_clean}
\end{equation}
where $\star_\perp$ denotes 2D correlation over the transverse grid $\mathcal{G}_{xy}$.

\subsection{Inverse problem formulation}

Given measured data $d \in D$, we estimate the object $\rho \in O$ by solving the regularized optimization problem
\begin{equation}
\hat{\rho}
=
\arg\min_{\rho \in O}
\left\{
f(\rho) + g(\rho)
\right\},
\label{eq:inverse_problem}
\end{equation}
where
\begin{equation}
f(\rho) \triangleq \frac{1}{2}\|H\rho-d\|_D^2
\label{eq:data_fidelity}
\end{equation}
is the data-fidelity term, and $g$ is a regularization functional that encodes constraints through indicator functions. In this work, we use
\begin{equation}
g(\rho)=\lambda\|\rho\|_1+\iota_{\mathbb{R}_+}(\rho),
\label{eq:g_definition}
\end{equation}
where $\lambda>0$ is the regularization parameter and $\iota_{\mathbb{R}_+}$ denotes the non-negativity indicator function, equal to $0$ when $\rho\ge 0$ and $+\infty$ otherwise. This formulation leads naturally to a proximal gradient method, here implemented using FISTA \cite{beck_fast_2009}.

The function $f$ is convex and differentiable, with gradient
\begin{equation}
\nabla f(\rho)=H^*(H\rho-d),
\label{eq:gradient_f}
\end{equation}
where $H^*$ denotes the adjoint of $H$ with respect to the inner products defined in \eqref{eq:euclidean_inner_products_clean}. The non-smooth term $g$, including both regularization and constraints, is handled through the proximal step. Evaluating the gradient in \eqref{eq:gradient_f} requires application of both forward $H$ and adjoint $H^*$ operators, and thus largely determines the computational cost of MB iterative reconstruction. Thanks to their depth-dependent transverse convolutional structure derived in the previous subsection, both operators admit efficient FFT-based implementations using 2D linear convolutions and correlations, with appropriate zero-padding and cropping to enforce linear rather than circular convolution. Additional linear operations arising from acquisition or preprocessing can be incorporated by composition within the same framework through a so-called measurement operator.

\section{Results and Discussion}
\subsection{Numerical validation of the forward model}

We first verify that the proposed FFT-based convolutional operator
produces \emph{discrete simulated data} identical to a direct numerical wave-equation propagation, under matched numerical
assumptions.
We compare the predicted measurements $d_{\mathrm{model}} = H\rho$ obtained with our implementation of
\eqref{eq:discrete_forward} against $d_{\mathrm{kW}}$ produced by k-Wave on the same discretization and acquisition geometry.
To ensure a fair comparison, the two simulations use identical boundary-condition settings. We compare the two forward models in the data domain using two error metrics:
(i) the relative $\ell_2$ error,
$\|d_{\mathrm{model}}-d_{\mathrm{kW}}\|_2/\|d_{\mathrm{kW}}\|_2$,
and (ii) the relative $\ell_\infty$ error,
$\|d_{\mathrm{model}}-d_{\mathrm{kW}}\|_\infty/\|d_{\mathrm{kW}}\|_\infty$. Over 100 test objects generated as Gaussian random initial-pressure distributions, both metrics remain at the level of numerical precision (see Table~\ref{tab:fwd_equiv}), confirming the numerical equivalence of the two approaches, provided that the depth-dependent transverse shift-invariant model is valid.

\begin{table}[ht]
\centering
\caption{\textbf{Forward model comparison.} Difference between the modelled acoustic data using either the proposed convolution-based operator or k-Wave implementation of pseudo-spectral numerical
acoustic wave propagation,
over 100 random test objects.}
\label{tab:fwd_equiv}
\begin{tabular}{lcc}
\toprule
Metric & Median & Max \\
\midrule
$\|d_{\mathrm{model}}-d_{\mathrm{kW}}\|_2/\|d_{\mathrm{kW}}\|_2$ 
& $2.19\times 10^{-15}$ & $2.38\times 10^{-15}$ \\
$\|d_{\mathrm{model}}-d_{\mathrm{kW}}\|_\infty/\|d_{\mathrm{kW}}\|_\infty$ 
& $2.35\times 10^{-15}$ & $3.43\times 10^{-15}$ \\
\bottomrule
\end{tabular}
\end{table}

\subsection{Computational complexity of the proposed approach}

We next assess the computational implications of replacing repeated numerical wave-field propagations with the proposed depth-dependent convolutional operator, which constitutes the central motivation of the present work. In the proposed formulation, each application of the forward operator $H$ (Eq.\eqref{eq:discrete_forward}) reduce to
batched \emph{two-dimensional} FFT-based convolutions in the transverse plane $(x,y)$, followed by depth-wise accumulation
and repeated over time samples.
Writing $N_{xy} = N_xN_y$, one application of the proposed forward operator consists of:
(i) $N_z$ 2D FFTs over $(x,y)$,
(ii) a Fourier-domain multiply-and-accumulate (MAC) over depth for each time sample, requiring
$\mathcal{O}(N_{xy}N_zN_t)$ complex MACs, and
(iii) $N_t$ 2D inverse FFTs over $(x,y)$.
The resulting per-application complexity is therefore
\begin{equation}
\mathcal{O}\!\left( (N_z + N_t)\,N_{xy}\log N_{xy} \;+\; N_{xy}N_zN_t \right).
\label{eq:complexity_ours}
\end{equation}
As correlations and convolutions can both be performed in the Fourier domain and share the same complexity, and as Nt and Nz play identical roles, the complexity of forward and adjoint operators computations is identical.
By contrast, k-Wave evaluates propagation using a pseudo-spectral time-domain scheme, where
spatial derivatives are computed using FFTs over the \emph{full three-dimensional} grid at every time step.
Denoting $N_{xyz}=N_xN_yN_z$ (again after any padding), a single k-Wave forward simulation has per-application complexity
on the order of
\begin{equation}
\mathcal{O}\!\left(N_t\,N_{xyz}\log N_{xyz}\right),
\label{eq:complexity_kwave}
\end{equation}
since each time step requires three-dimensional FFT-based evaluations of spatial derivatives
(and associated pointwise operations) over the full volume, which dominate the runtime in 3D in practice~\cite{treeby_k-wave_2010}. The proposed convolutional operator requires to propagate the acoustic field over the entire volume grid once using k-Wave and store a library of depth-indexed impulse responses $\{h_z\}_{z\in\mathcal{G}_z}$. We can then apply
$H$ and $H^*$ repeatedly during reconstruction following equations \eqref{eq:discrete_forward} and \eqref{eq:adjoint_final_clean} in the Fourier domain, that is without any further wave-equation solving.
A direct construction of this library would require running k-Wave separately for each depth by placing a unit-amplitude
point absorber centred in the transverse plane and recording its response on the sensor plane, leading to $N_z$ 3D propagations.
However, we simplify this further by exploiting acoustic reciprocity to generate the full library from a \emph{single} propagation. A point source is placed at the bottom of the grid, and resulting time series are recorded on virtual sensors defined across all depth planes. The depth indexed impulse responses are then obtained by reshaping these recorded fields.
This yields the same impulse-response library that would be obtained by repeating point-source simulations at each depth, while reducing the
pre-computation cost from $N_z$ wave-equation simulations to one.

\begin{figure}[ht]
    \centering
    \includegraphics[width=0.7\linewidth]{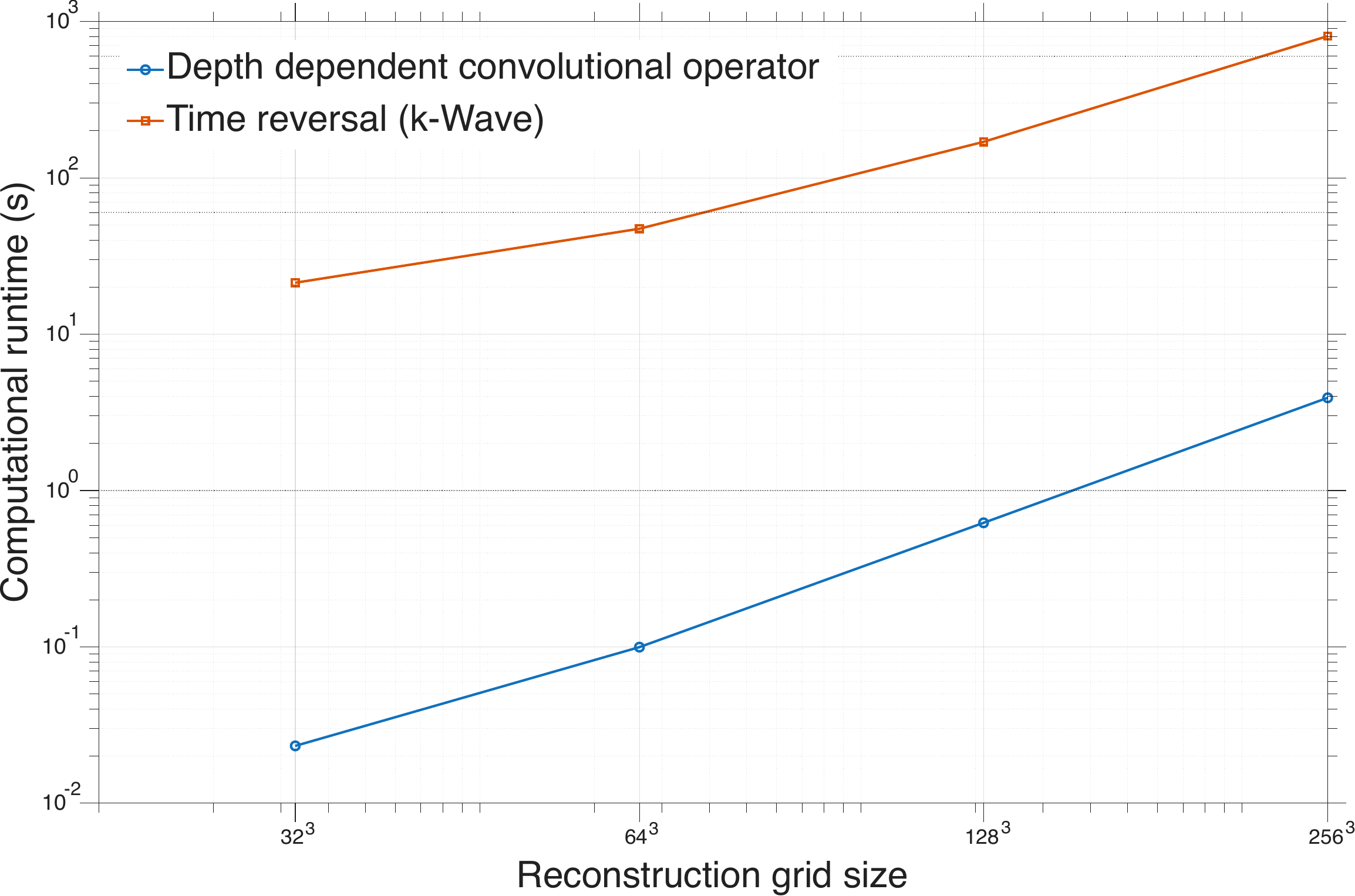}
    \caption{\textbf{Forward operators runtime as a function of grid size.}
    Single forward operator runtime comparison between k-Wave implementation of pseudo-spectral numerical acoustic wave propagation and the proposed depth-dependent convolutional operator with $N_t=1000$ time steps, as a function of the 3D grid size. 
    }
    \label{fig:time_comp}
\end{figure}

Benchmarks shown in Figure~\ref{fig:time_comp} were performed in MATLAB on a Linux workstation (SUSE Linux Enterprise Server 15 SP4, kernel 5.14.21) equipped with dual AMD EPYC 7662 CPUs
(2$\times$64 cores), 2.0~TiB RAM, and an NVIDIA 16~GB Tesla T4 GPU. For each grid size and each method, runtimes were measured over three repeated runs and we report the mean computational time. For a like-for-like comparison, we report k-Wave runtimes using its CPU implementation and disabled on-screen plotting to avoid visualization overhead,
following the benchmarking practice discussed in~\cite{treeby_k-wave_2010}. The proposed operator is also CPU-based here and is naturally amenable to GPU acceleration because its core computations are batched 2D FFTs and pointwise complex multiplications. Across the tested 3D grid sizes, the proposed forward evaluation is consistently \emph{two to three orders of magnitude} faster than k-Wave.
For example, for a $256^3$ grid, k-Wave requires on the order of $10^3$ seconds per forward evaluation, whereas the convolutional operator evaluates in a few seconds.
Since the adjoint operator $H^*$ is implemented with the same computational structure (transverse correlations evaluated in the Fourier domain, and summation over time and depth),
the same acceleration applies to adjoint evaluations.
As the runtime of each iteration is dominated by the applications of $H$ and $H^*$, this leads to a similar acceleration of the entire iterative reconstruction process.

\subsection{Image reconstruction of phantoms}
To assess the validity of the assumptions required for our the proposed depth-dependent convolutional forward model,  we iteratively reconstruct images of phantom samples from experimental photoacoustic measurements acquired with a planar Fabry--P\'erot (FP) sensor,
following the acquisition and preprocessing protocol in~\cite{saucourt_fast_2023}.
Two representative phantom configurations are considered:
(i) near point-like absorbers (black polyethylene beads, diameter 10--20~$\mu$m) and
(ii) extended, vessel-like structures (black nylon wires, nominal diameter 20~$\mu$m), embedded in agarose and immersed in water. Prior to reconstruction, the recorded time series were bandpass filtered to approximately match the dominant spectral content of each phantom:
10--120~MHz for the wire dataset and 50--120~MHz for the bead dataset.

All reconstructions are computed from the same datasets using three approaches:
(i) conventional TR reconstruction, as implemented in k-Wave, (ii) direct \emph{adjoint reconstruction} under the proposed discrete convolutional model, defined by applying the discrete adjoint $H^*$ to the measured data,
and (iii) iterative reconstruction using the same forward-adjoint pair within the previously described constrained optimization framework (FISTA with non-negativity and $\ell_1$ regularization), with $\lambda = 2\times10^{-5}$ for the \textit{phantoms}. For both datasets, measurements were acquired on a $101 \times 101$ scan grid with a pitch of $20~\mu\mathrm{m}$, corresponding to a field of view of $2.02 \times 2.02~\mathrm{mm}^2$. Images were then reconstructed on an upsampled grid with isotropic voxel size $10~\mu\mathrm{m}$.
Reconstruction was restricted to the depth range containing the target structures.
For the wire dataset, the effective reconstruction volume used in the forward model was $202 \times 202 \times 17$ voxels, corresponding to a spatial extent of $2.02 \times 2.02 \times 0.17~\mathrm{mm}^3$.
For the bead dataset, the corresponding volume was $202 \times 202 \times 37$ voxels, i.e.\ $2.02 \times 2.02 \times 0.37~\mathrm{mm}^3$. 

In standard k-Wave-based pipelines, such a mismatch between measurement and reconstruction grids is often handled by interpolating the measured waveforms before TR. In contrast, our formulation represents the acquisition geometry explicitly through a discrete sampling operator in the forward model, which maps object estimates defined on the refined reconstruction grid to predicted measurements on the coarser scan grid; its adjoint maps residuals back to the reconstruction grid, ensuring discrete consistency without waveform interpolation.

\begin{figure}[ht]
    \centering
    \includegraphics[width=.8\linewidth]{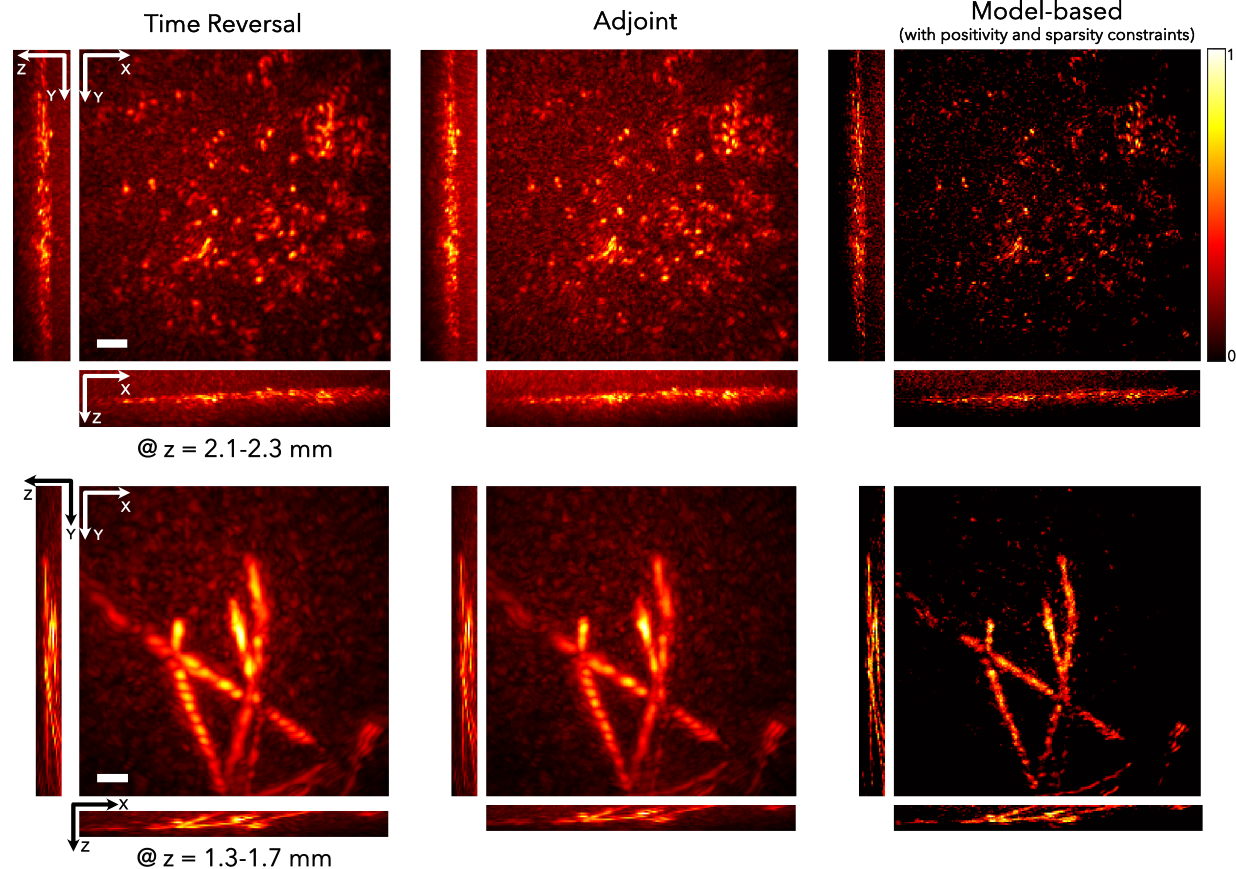}
    \caption{\textbf{Image reconstructions of phantoms.}
    Image reconstructions from experimental datasets, comparing direct baselines and iterative MB approach: (left) k-Wave TR, (middle) adjoint reconstruction under the proposed depth-dependent convolutional operator,
    (right) FISTA reconstruction with non-negativity and $\ell_1$ regularization ($\lambda = 2\times10^{-5}$, 15 iterations), using the proposed forward and adjoint operators.
    Two representative phantom samples are tested: a sparse collection of 10-20 µm black beads (top) and elongated structures formed by a 20 µm black wire, mimicking a vessel-like organisation (bottom). Scale bars: $200~\mu$m.
    }
    \label{fig:exp_results}
\end{figure}

In general, TR and direct adjoint reconstruction are distinct direct operators in photoacoustic imaging~\cite{arridge_adjoint_2016}.
TR enforces time-reversed measurements through boundary conditions, whereas adjoint reconstruction corresponds to the discrete transpose mapping associated with the chosen forward model and inner product.
Accordingly, small differences between TR and adjoint reconstructions are expected under matched assumptions, while the overall spatial agreement in Fig.~\ref{fig:exp_results} confirms consistent propagation modelling. As expected \cite{dean-ben_practical_2022}, iterative FISTA reconstruction improves contrast and reduce background artefacts relative to both direct reconstruction baselines. These improvements arise from explicitly enforcing data fidelity together with physical and statistical priors (non-negativity and sparsity), rather than from any modification of the forward model. Crucially, the large reduction in per-application runtime provided by the proposed operator makes such iterative, constraint-aware reconstructions practical for large 3D experimental datasets.

\subsection{Image reconstruction of \textit{in vivo} vasculature}

To demonstrate the performance of the proposed model and the validity of the underlying assumptions \emph{in vivo}, we reconstructed images from a dataset acquired on a human forearm. For this experiment, a commercial all-optical photoacoustic imaging system with a planar FP sensor was used (LightEcho-RH, deepColor Imaging), similar to the system described in \cite{huynh_fast_2024}. In this system, 32 interrogations spots were simultaneously scanned over the sensor, and corresponding time series were recorded at each location. A total field of view of $19.2 \times 16.3~\mathrm{mm}^2$ was scanned with a pitch of $50~\mu\mathrm{m}$ in both transverse directions, corresponding to a $384 \times 326$ XY transverse grid.
Standard preprocessing steps were applied prior to reconstruction, including removal of DC offsets and bandpass filtering (1--25~MHz) to suppress low-frequency drift and high-frequency noise. Reconstruction was performed without any upsampling in the transverse plane in this case, using an isotropic voxel size of $50~\mu\mathrm{m}$. The effective grid used in the forward model was $384 \times 326 \times 96$ voxels, corresponding to a spatial extent of $19.2 \times 16.3 \times 4.8~\mathrm{mm}^3$.

\begin{figure}[ht]
    \centering
    \includegraphics[width=.7\linewidth]{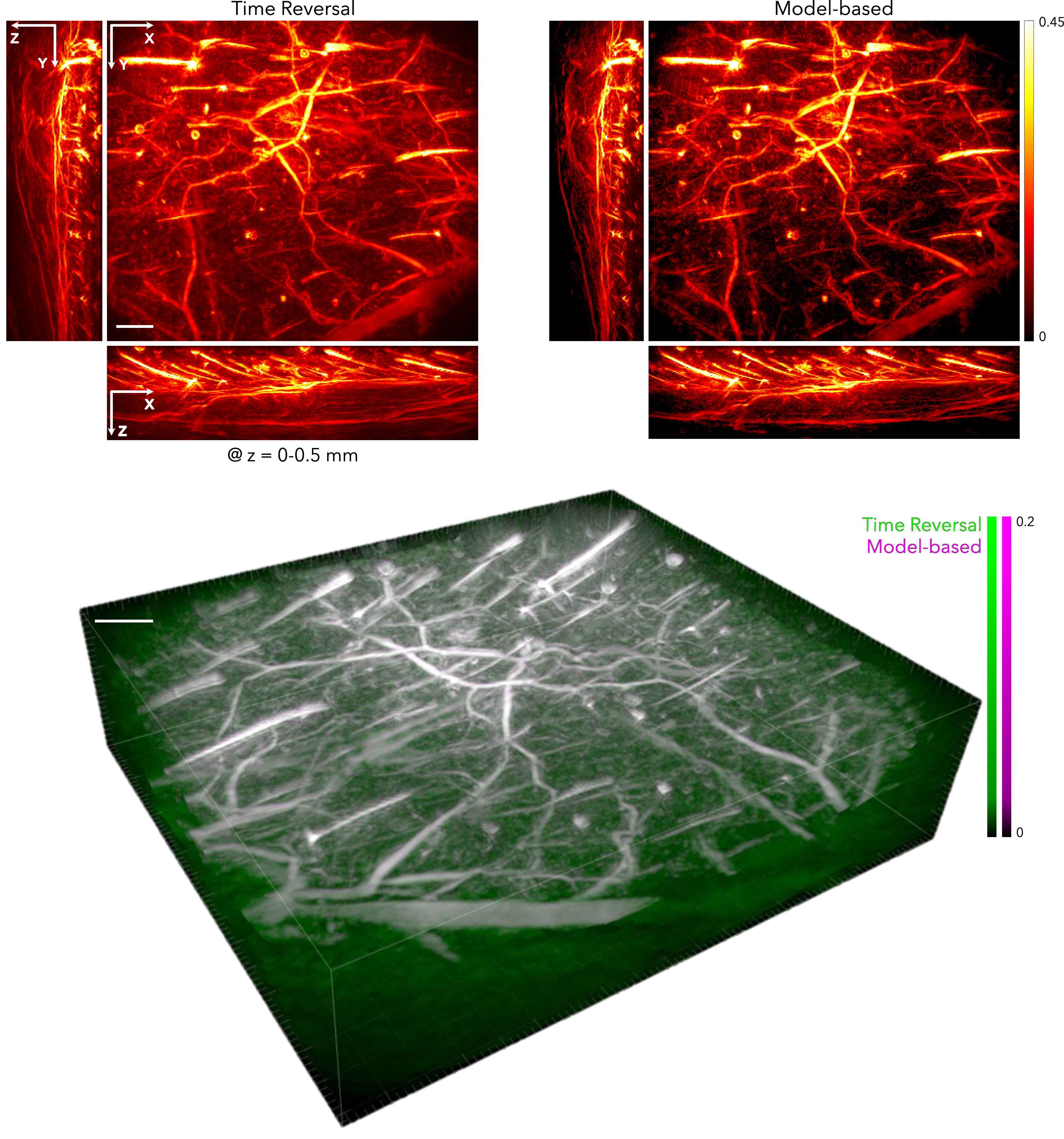}
    \caption{\textbf{Image reconstructions of in-vivo human forearm vasculature.}
    Comparison of TR and iterative MB (FISTA with non-negativity and $\ell_1$ regularization) reconstructions of an \textit{in vivo} human forearm. Overlay at the bottom highlights the higher noise level (green background) in TR reconstruction, while objects features appear in both reconstructions (white structures).  Scale bars: 2 mm.
    }
    \label{fig:invivo_results}
\end{figure}

Figure~\ref{fig:invivo_results} compares the TR reconstruction with the iterative reconstruction obtained using the proposed forward--adjoint pair and FISTA with positivity and $\ell_1$ regularization ($\lambda = 5\times10^{-5}$, 15 iterations). Both reconstructions recover the main vascular organization of the human forearm and show consistent anatomical continuity. As expected, structures common to both reconstructions remain spatially consistent, and the MB reconstruction significantly reduces background noise and improves the delineation of finer vessels. This effect is particularly apparent in the 3D overlay rendering. These observations are consistent with the phantom experiments, and indicate that the proposed depth-dependent convolutional model remains accurate for \textit{in vivo} scenarios.

\section{Discussion}

The aim of this work was to accelerate iterative MB reconstruction in 3D photoacoustic tomography with planar sensors, without stepping away from a physically consistent acoustic description. Rather than numerically propagating the acoustic field anew at each iteration, we exploit the transverse shift invariance of this acoustic propagation along the sensor plane. In this framework, the forward and adjoint operators can be evaluated efficiently using FFT-based convolutions from a precomputed library of depth-dependent transverse shift invariant impulse responses. As a result, the dominant computational burden associated with iterative reconstruction was reduced by more than two orders of magnitude while still yielding accurate reconstructions, including \textit{in vivo}.

Besides curing the ill-posed nature of acoustic inversion, MB reconstruction approaches play a central role when considering complex acquisition strategies. In particular, compressed sensing or other non-standard measurement strategies can benefit from our approach by composing the physical model with an appropriate measurement operator and its adjoint.

This gain, however, comes with a clear trade-off. The proposed method relies on computing and storing a large structured operator, which in 3D can result in substantial memory requirements. Here, the stored precomputed operator used during iterative reconstruction occupied approximately 17.3~GiB for the wire phantom dataset and 78.1~GiB for the \textit{in vivo} case, both in single-precision complex format. In other words, the method exchanges computation time for memory. Although the proposed method requires storing a large precomputed operator, its memory footprint remains far smaller than that of a full 3D system matrix. This reduction is achieved by exploiting transverse shift invariance, so that only depth-indexed impulse responses need to be stored instead of voxel-specific responses throughout the volume. The memory cost is therefore not eliminated, but compressed into a form that remains tractable for realistic 3D reconstructions. Another limitation is that the transverse shift invariance assumption underlies homogeneous acoustic properties throughout the reconstructed volume, as well as identical element impulse response throughout the sensor, which is not strictly met in practice. Yet this approach turned out to be valid for widely used ultrasound sensors, as shown in both phantom and \textit{in vivo} experiments. These limitations suggest natural directions for future work rather than fundamental obstacles. On the modelling side, the present framework could be extended by composing the structured forward model with additional linear measurement operators to account for acquisition-dependent effects or to partially compensate for departures from the idealized assumptions. On the implementation side, the forward and adjoint operators are highly structured and therefore well suited to GPU acceleration, which could further reduce reconstruction times and improve scalability.

\section*{Funding}
This work was supported by the European Research Council (Starting Grant ALPINE, 101117471) and by the French National Research Agency through the ``Investissements d'Avenir'' programme (ANR-21-ESRE-0003).

\section*{Acknowledgments}
We thank Ben Cox for fruitful discussions.

\bibliographystyle{ieeetr}
\bibliography{references}

\end{document}